# Magnetic Spring Device

A. B. Hassam and J. C. Rodgers
*Institute for Research in Electronics and Applied Physics*
*University of Maryland, College Park*

**Abstract**
A cylindrical system is proposed that will store magnetic energy in a localized azimuthal field that can then be quickly released on Alfvenic timescales, accompanied by the formation of a flowing Z-pinch plasma. The magnetized plasma is MHD in character and will have unilateral axial momentum with Alfvenic speeds. Conventional plasma gun injectors (Marshall type) have a limited parameter space of operation. The "magnetic spring" momentum injector differs from Marshall guns in that it has an already stored strong magnetic field before release. The resulting parameter space is much broader. There are possible applications to momentum injectors for fusion and to plasma and rail guns.

Marshall guns[1,2] accelerate magnetized plasma by quickly creating a magnetic field and plasma in a localized region such that the unbalanced magnetic pressure expands outward carrying frozen-in plasma with it. The initial magnetic field is created by an electric discharge between electrodes initiated by a Paschen breakdown due to an applied voltage.

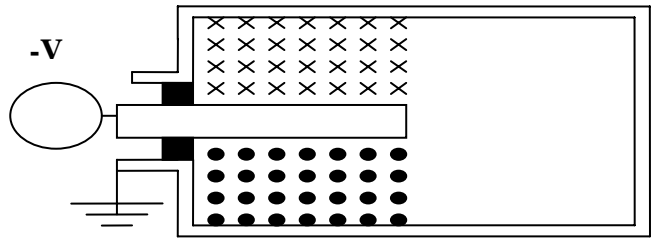

Fig. 1

A typical setup is shown in Fig 1: A high voltage conductor is wrapped around and spaced off of the main cylinder by two wrap-around insulators (shaded dark). The high voltage is applied in the presence of a fill gas, initiating a Paschen breakdown. The resulting radial current creates an azimuthal B field in the chamber between the electrodes and the resulting **j x B** force accelerates the field and embedded plasma rightward, axially into the main chamber.

This method of acceleration has a limited parameter space as shown in Fig 2. Here, the ordinate is the fill pressure density, N (eventually ionized to plasma density, n) and the abscissa is the maximum magnetic field that can be generated by the Marshall gun at the end of the breakdown phase. The green shaded region is the accessible part of the n-B parameter space, explainable as follows: The two horizontal lines represent the maximum and minimum fill pressures between which one would obtain Paschen breakdown for a given voltage. The linear in B line is the line $\omega_{ce} = \nu_{eN}$ (in terms of N ~ n and B, this becomes n = const x B; $\omega_{ce}$ is the electron cyclotron frequency, $\nu_{eN}$ is the electron-neutral frequency): to run a significant current across a magnetic field, the electron-neutral mean free path has to be smaller than the electron Larmor radius (in the self-consistent B field created by the current), ie, $\omega_{ce} < \nu_{eN}$. If the B field becomes too strong, ie, $\omega_{ce} > \nu_{eN}$, the current is clamped as, beyond this, the plasma resistance becomes too high. Thus, the maximum field of operation is determined by $\omega_{ce} < \nu_{eN}$.

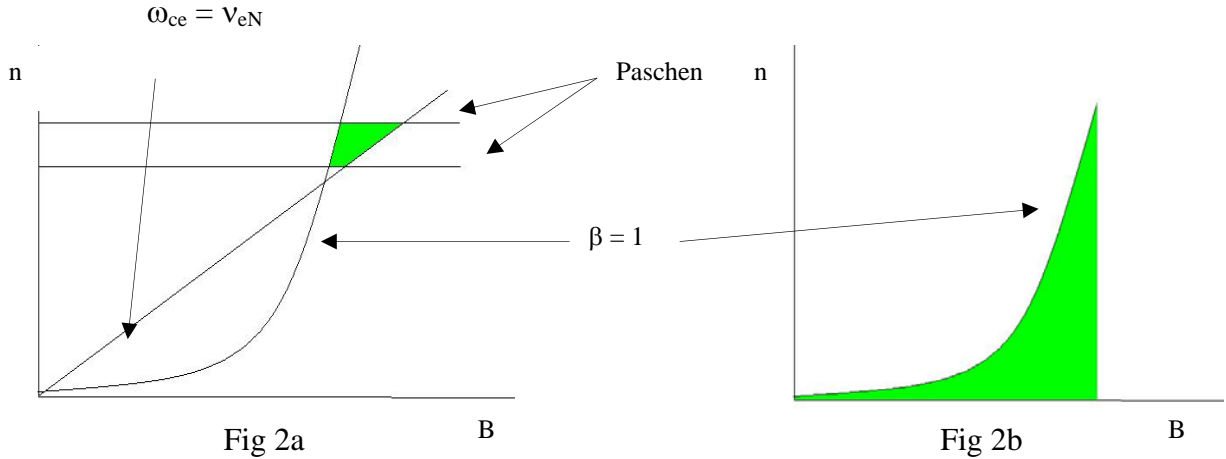

Fig 2a   Fig 2b

Finally, the line proportional to $B^2$ is the $\beta = 1$ line, where $\beta$ is the ratio of thermal to magnetic pressure. This is not a limitation on formation, per se; rather, operation at $\beta < 1$ would be preferred as the stronger magnetic field would then provide better heat insulation for the plasma; else, the plasma expands and cools off. All told, the three boundaries lead to the (schematic) green shaded region as being accessible.

In this note, we propose an alternative method for plasma acceleration. This method, shown in Fig. 3, is almost identical to that in the previous Figure, except for the double solid line joining the outer electrode and the inner vessel. This double solid line represents a thin, azimuthal metallic foil. In this system, the applied voltage will make a current flow thru the foil (with no plasma breakdown for low enough fill pressures and, so, no plasma current) and create a vacuum magnetic field as strong as is desired. The foil thickness is now calibrated such that, at the peak of the field strength, the magnetic pressure is just sufficient to overcome the tensile strength of the foil (alternatively, the foil will melt); such a "fuse" will cause a "blowout" of the vacuum magnetic energy into the main chamber, thus interrupting the current. Of course, Lenz' Law will try to prevent such a vacuum blowout and current will persist initially in the plasma formed from the destroyed foil. Very quickly, however, the background fill gas will ionize, from the strong inductive E fields generated, creating, in eventuality, an MHD-equilibrated Z-pinch plasma flowing axially into the main chamber.

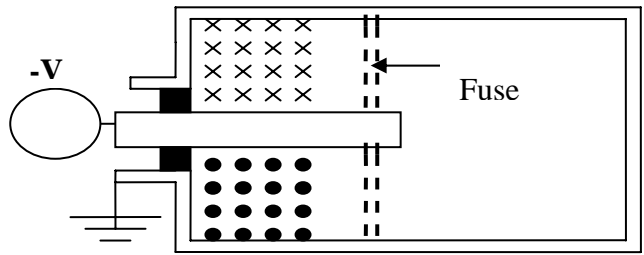

Fig. 3

The "fuse" method ("Magnetic Spring") is expected to be operative over a much wider area in the experimental parameter space.   Fig. 2b is the parameter space of operation: this method does not rely on a Paschen breakdown, thus the horizontal lines are not relevant.  The maximum B that can be attained by this method is independent of the fill pressure and thus the linear in B line is not relevant.   Since we desire $\beta < 1$, the green shaded area is the operating space, significantly broader than the Method 1 space.

The Magnetic Spring system has a low duty factor, would have to be reset shot by shot.  An enhanced duty cycle experiment can be imagined.  A schematic is displayed in Fig. 4.   In this case there is a spring-loaded "slider" that would function as an opening switch, initiated by a critical magnetic pressure

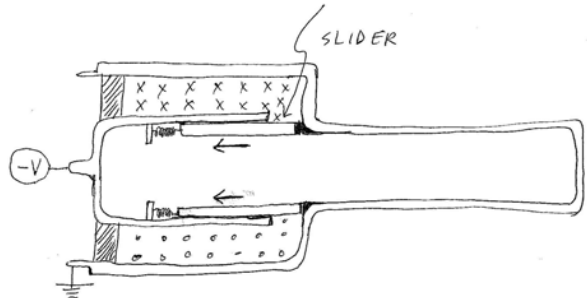

Fig. 4

[1] J. Marshall, *Phys Fluids* **3**, 134 (1960)
[2] C. T. Chang, *Phys Fluids* **4**, 1085 (1961)